\documentclass{aa}  
\usepackage{color}
\bibpunct{(}{)}{;}{a}{}{,}
\usepackage{graphicx}
\usepackage{txfonts}
\usepackage{lipsum}
\usepackage{subcaption}         
\usepackage{lscape}             
\usepackage{orcidlink}
\DeclareUnicodeCharacter{FF0C}{,} 
\usepackage{xcolor}
\usepackage{placeins}          
\usepackage{threeparttable}
\makeatletter
\let\@ln@colored\@empty
\let\@ln@normal\@empty
\renewcommand{\linenumberfont}{}
\let\linenumbers\relax
\let\nolinenumbers\relax
\@ifpackageloaded{linenoaa}{%
    \PackageWarningNoLine{aa}{linenoaa package loaded but disabled}%
    \let\linenumbers\relax
    \let\nolinenumbers\relax
}{}
\makeatother          
\usepackage{hyperref}

\begin{document}

   \title{The X-ray properties and structure of A3571 up to R$_{500}$}

\author{%
    {X. Zheng}\inst{1,2}%
  \and {H. Yu}\inst{1}%
  \and {S. Jia}\inst{2}%
  \and {C. Li}\inst{2}%
  \and {X. Hou}\inst{1}%
  \and {A. Liu}\inst{1,3,4}%
  \and {Y. Chen}\inst{2}%
  \and {\\H. Feng}\inst{2}%
  \and {L. Song}\inst{2}%
  \and {C. Liu}\inst{2}%
  \and {F. Lu}\inst{2}%
  \and {S. Zhang}\inst{2}%
  \and {W. Yuan}\inst{5}%
  \and {J. Sanders}\inst{3}%
  \and {\\J. Wang}\inst{6}%
  \and {K. Nandra}\inst{3}%
  \and {W. Cui}\inst{2}%
  \and {J. Guan}\inst{2}%
  \and {D. Han}\inst{2}%
  \and {C. Jin}\inst{4,5,7}%
  \and {Y. Liu}\inst{5}%
  \and {\\J. Xu}\inst{2}%
  \and {J. Zhang}\inst{2}%
  \and {H. Zhao}\inst{2}%
  \and {X. Zhao}\inst{2}%
}

\authorrunning{X.Y. Zheng et al.}

\institute{%
    School of Physics and Astronomy, Beijing Normal University, Beijing 100875, China\\
    \email{yuheng@bnu.edu.cn}
  \and State Key Laboratory of Particle Astrophysics, Institute of High Energy Physics, Chinese Academy of Sciences, Beijing 100049, China \\
  \email{jiasm@ihep.ac.cn}
  \and Max Planck Institute for Extraterrestrial Physics, Giessenbachstrasse 1, 85748 Garching, Germany
  \and Institute for Frontiers in Astronomy and Astrophysics, Beijing Normal University, Beijing 102206, China
  \and National Astronomical Observatories, Chinese Academy of Sciences, 20A Datun Road, Beijing 100101, China
  \and Department of Astronomy, Xiamen University, Xiamen 361005, China
  \and School of Astronomy and Space Science, University of Chinese Academy of Sciences, 19A Yuquan Road, Beijing 100049, China
}

   \date{Received September 30, 20XX}

  \abstract
{Abell 3571 is a nearby, X-ray-bright galaxy cluster located in the Shapley Supercluster. Although it appears morphologically relaxed in X-ray images, multiwavelength observations reveal subtle indications of residual dynamical activity likely associated with past merger events. Using wide-field ($1^{\circ} \times 1^{\circ}$) data from the Einstein Probe Follow-up X-ray Telescope (\textit{EP-FXT}), we extended measurements of the cluster’s properties beyond its $R_{500}$ radius.
We detect surface-brightness excesses on both the northern and southern sides, consistent with previous \textit{XMM-Newton} results. The temperature, pressure, and entropy in the northern excess region are lower than the average values, whereas those on the southern side are slightly higher. However, we find no evidence of cold fronts or shocks. These features can be interpreted as sloshing motions triggered by an off-center minor merger. Our findings suggest that, despite its symmetric appearance, A3571 is still recovering from a minor merger and is currently in a post-merger phase. This work also demonstrates the excellent capability of \textit{EP-FXT} for probing the outskirts of galaxy clusters.}


   \maketitle

\section{Introduction}\label{sec:intro}

Galaxy clusters are the most massive gravitationally bound systems in the Universe. The majority of their baryonic matter exists in the form of a hot, diffuse intracluster medium (ICM), which emits strongly in the X-ray band \citep{kravtsov2012formation,vikhlinin2014clusters}. Although many clusters appear to be approximately relaxed and in equilibrium in observations, their formation is fundamentally a highly dynamic process. 
Frequent merger events and continuous matter accretion produce pronounced dynamical effects in the ICM, driving significant departures from hydrostatic and thermal equilibrium \citep{markevitch2007shocks,kravtsov2012formation,2022A1914,Liu2018}. Understanding these disturbances is therefore essential for constraining the physical processes that shape the formation and evolution of galaxy clusters, as well as for providing key observational insights into the assembly of large-scale cosmic structures.

X-ray observations have revealed the characteristic signatures of such dynamical activity, confirming the widespread presence of nonequilibrium structures in galaxy clusters, among which shocks and cold fronts are the most representative. Shocks, typically generated by high-velocity mergers, heat the ICM on short timescales and appear in X-ray images as sharp, arc-shaped edges \citep{markevitch2007shocks}. The Bullet Cluster \citep{2008bullet,2002ApJMarkevitchbullet} and Abell 3667 \citep{2009ApJ3667} provide two of the most striking examples. Cold fronts, by contrast, appear as contact discontinuities where denser but cooler gas is juxtaposed against hotter surroundings \citep{2000ApJMarkevitchcf,2001ApJvik3667,markevitch2007shocks}. They are generally linked to the displacement of cool core gas induced by gravitational perturbations, such as those triggered by minor mergers \citep{Ascasibar2006,roediger2011gas,zuhone2010stirring}. In relaxed clusters, such perturbations can drive sloshing motions of the central gas, producing multiple cold fronts at different radii. Over time, these sloshing motions can evolve into intricate spiral structures, which are observed as concentric edges in X-ray images \citep{sonkamble2024cool,kadam2024sloshingA795,kadam2024sloshingA2566}. To date, spiral-like cold front structures have been identified in a number of well-studied clusters, including Abell 496 \citep{2014A&Aghizza496}, Abell 795 \citep{kadam2024sloshingA795}, Perseus \citep{2017MNwalkerpers,2006MNfabianpers}, and Virgo \citep{roediger2011gas,2025A&Azheng}.

Abell 3571 (A3571) is a nearby rich galaxy cluster at redshift 0.039 \citep{2010A&Arosset}, located in the Shapley Supercluster \citep{shapley1991}.
The brightest cluster galaxy (BCG) of A3571 is MCG--05--33--002,
which exhibits a pronounced north--south elongation extending over several hundred kiloparsecs \citep{kemp19910}.
Although A3571 is classified as a relaxed cool-core cluster, \citet{2019lagan} found no evidence of a central temperature drop in its temperature map.
In addition, \citet{2005xmm3571} reported a noticeable north--south temperature asymmetry, suggesting that A3571 recently experienced -- or is still experiencing -- dynamical evolution.

Early X-ray observations revealed that the ICM of A3571 exhibited an average temperature of $7.6_{-0.9}^{+1.2}$ keV, based on \textit{EXOSAT}\footnote{The European X-ray Observatory Satellite}data \citep{1990edge}, or $6.9 \pm 0.2$ keV, based on \textit{ASCA}\footnote{The Advanced Satellite for Cosmology and Astrophysics} data \citep[the radial temperature profile extends to $30^{\prime}$ at \textit{ASCA}'s $\sim$ $3^{\prime}$ half-power diameter;][]{asca1998}, with a central metal abundance reaching as high as 0.6 times the solar value at R~\textless~$0.5^{\prime}$ and declining to about 0.3 at a radius of $7^{\prime}$. \textit{XMM-Newton} radial profiles further trace the temperature and metallicity distributions out to approximately $10^{\prime}$ \citep{2005xmm3571}.
\textit{ROSAT}\footnote{The ROentgen SATellite} imaging revealed a nearly spherical X-ray brightness distribution, with the intracluster gas appearing azimuthally symmetric and exhibiting only slight ellipticity \citep{3571ROSAT&ASCA}.
Based on high-resolution observations from \textit{XMM-Newton}, the central region of  A3571 shows pronounced temperature and metallicity inhomogeneities, which appear as a patchy distribution. A temperature difference of up to 2 keV is detected between the northern and southern regions, suggesting that the cluster has undergone complex dynamical processes, such as the infall of subgroups \citep{2005xmm3571}.

Studies of A3571 not only examined the properties of the hot gas but also attempted to infer the gravitational mass and dark matter distribution within the cluster.  
\citet{3571ROSAT&ASCA} combined \textit{ASCA} temperature data with \textit{ROSAT} imaging data to estimated the total mass of the cluster to be approximately $M_{500}=7.8_{-2.2}^{+1.4} \times 10^{14}\ h_{50}^{-1}\ M_{\odot}$ with $R_{500}=1.7~h_{50}^{-1}$~Mpc at 90\% confidence, which means the cluster is in hydrostatic equilibrium. 
Subsequently, \citet{nevalainen2001} derived the total mass of A3571, \( M(<r_{178}) = 9.1 \times 10^{14}\ h_{50}^{-1}\ M_{\odot} \), by combining the \textit{Beppo}SAX and \textit{ASCA} temperature profiles with a dark matter density model under the assumption of hydrostatic equilibrium. The corresponding gas mass fraction was estimated to be \( f_{\mathrm{gas}} = 0.26^{+0.05}_{-0.10} \times h_{50}^{-3/2} \).

Although X-ray observations indicate that A3571 exhibits an overall spherically symmetric structure with a central cooling flow, multiwavelength data have revealed a more complex dynamical history. Early studies found that the spatial distribution of galaxies and the pronounced north–south elongation of the central cD galaxy are prominent characteristics \citep{kemp19910}, possibly suggesting that the cluster underwent merger or disturbance events. 
Subsequently, spectroscopic analyses revealed an asymmetric radial velocity distribution among member galaxies and identified multiple velocity substructures, further supporting an evolutionary scenario in which several smaller galaxy groups successively merged into the main cluster \citep{quintana1993spectroscopic}. 
\citet{2002venturi} conducted a radio analysis of the galaxy cluster complex consisting of A3571, A3572, and A3575, and concluded that the system is in a late merger stage, with its radio properties reflecting a relatively relaxed dynamical state. 
More recent multiwavelength observations have not detected large-scale radio halos or significant active galactic nucleus activity, indicating that A3571 is currently in a relatively relaxed, post-merger evolutionary stage.

Early multiwavelength observations suggest that A3571 is in a relatively relaxed, late post-merger stage. However, our understanding of its merger history remains limited, particularly regarding the hot gas in the outskirts. Therefore, we aim to conduct a more systematic and in-depth investigation of the gas distribution and merger state of A3571 using large field-of-view (FoV) X-ray imaging observations.

The Follow-up X-ray Telescope (FXT) is one of the key payloads aboard the \textit{Einstein} Probe (EP) mission, which was launched on January 9, 2024 \citep{yuan2022einstein,Yuan2025}. It features a FoV of approximately 1 square degree, an angular resolution of $9.6^{\prime \prime}$ per pixel, and a half-power diameter  spatial resolution of $22^{\prime \prime}$, with an energy band of 0.3-10 keV.
The \textit{EP-FXT} also benefits from an exceptionally low particle background, making it well suited for detecting faint diffuse X-ray emission.

This paper is organized as follows: In Sect. \ref{sec:data} we describe the data used in our analysis. In Sect. \ref{sec:3} the analysis of the X-ray and optical observations is described. In Sect. \ref{sec:discussion} we discuss cluster dynamics and possible subclusters. We summarize our results in Sect. \ref{sec:conculsion}.
We adopt $H_{0}=(67.4\pm0.5)~\mathrm{km\,s^{-1}\,Mpc^{-1}}$ and\ $\Omega_{m}=0.315\pm0.007$ \citep{2020plank}, and the best-fit parameters from the spectral fitting are reported with 90\% confidence intervals.

\section{Observations and data reduction} \label{sec:data}
\subsection{X-ray data}

During the performance verification phase of the EP mission, A3571 was observed with the \textit{EP-FXT} in full-frame (FF) mode from March 30 to 31, 2024 (ObsID 11904194436), with the total effective exposure time of 40ks.
The data reduction followed standard procedures using the FXT Data Analysis Software (\texttt{FXTDAS}, version 1.20)\footnote{\url{http://epfxt.ihep.ac.cn/analysis}} \citep{zhaohs2025}, 
with additional support from general analysis tools provided by \texttt{CIAO} (version 4.16)\footnote{\url{https://cxc.cfa.harvard.edu/ciao/}} and \texttt{HEASoft} (version 6.33.1)\footnote{\url{https://heasarc.gsfc.nasa.gov/docs/software/lheasoft}}.

\begin{figure}[h]
\centering
\includegraphics[width=\linewidth]{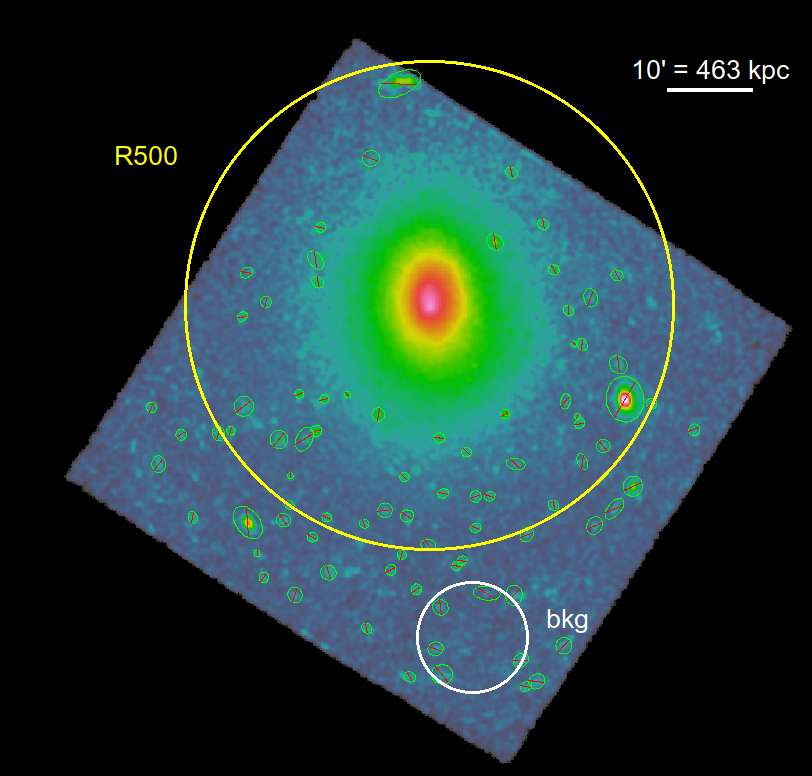}
\caption{Particle-background-subtracted, vignetting-corrected, and smoothed \textit{EP-FXT} image of A3571 in the 0.3–7.0 keV energy band. The green circle is $R_{500}$ (1.29 Mpc).} 
\label{fig:image}
\end{figure}

We used the \texttt{fxtchain} pipeline to generate a series of calibrated event files for subsequent scientific analysis. 
To eliminate the effect of out-of-time  events, which account for approximately 0.23\% of the total detected events in the FF mode, simulated out-of-time  event files were created using \texttt{fxtootest} and were subsequently subtracted from the original images using \texttt{ftimgcalc}.
Since the particle background is uniformly distributed and not affected by the telescope’s vignetting function, it was removed before applying vignetting correction. The background spectrum was extracted from the storage area of the event file and modeled using \texttt{fxtbkggen} \citep{zhang2025fxt}, after which it was subtracted from the image.
Vignetting, caused by the decrease in effective area at larger off-axis angles, leads to reduced photon flux at the edges of the FoV. To correct for this effect and improve the S/N, we used \texttt{reproject\_image} from CIAO to co-align and merge the images and vignetting-corrected exposure maps from FXT-A and FXT-B. The merged images were then divided by the combined exposure map using \texttt{farith} to produce a vignetting-corrected image.
After vignetting correction, point sources were identified using the \texttt{wavdetect} from CIAO and excluded to avoid contamination in the analysis of diffuse emission.
Subsequently, the smooth were conducted with a $14.4^{\prime\prime} ~\sigma$ Gaussian kernel.
The final X-ray image in the 0.3–7.0 keV band is presented in Fig.~\ref{fig:image}.

To compare \textit{EP-FXT} with \textit{XMM-Newton}, we selected the European Photon Imaging Camera (EPIC) MOS1 observation of A3571 conducted by \textit{XMM-Newton} in July 2002 (ObsID 0086950201, with an exposure time of 33 ks) and processed the data using the \textit{XMM-Newton} Science Analysis System (\texttt{SAS}\footnote{\url{https://www.cosmos.esa.int/web/xmm-newton/sas}}) in FF mode.
Time intervals affected by flares were removed by screening the high-energy light curve.

\subsection{Optical data}

The photometric catalog used in this study was derived from the Dark Energy Spectroscopic Instrument (DESI) Legacy Imaging Surveys Data Release 10 \citep{dey2019}. We selected a $2.0^{\circ} \times 1.5^{\circ}$ field encompassing the A3571 cluster and its surrounding region, spanning $205.87^{\circ} < \mathrm{R.A.} < 207.87^{\circ}$ and $-33.8^{\circ} < \mathrm{Dec} < -32.3^{\circ}$, with a  magnitude limit of mag$_r$ $<$ 22 . Cluster member candidates are identified using the red sequence method, by selecting galaxies within 3$\sigma$ of the best-fit red sequence and further restricting their photometric redshifts to the range $0.019 \leq z \leq 0.079$. This yields a sample of 399 galaxies.

To examine the spatial distribution of galaxies in and around A3571, we generated a smoothed luminosity density map based on this photometric sample. Following the method of \citet{2013MNwen}, the \textit{r}-band luminosity of each galaxy was estimated using its redshift and apparent magnitude. These luminosities were then mapped onto a 200 $\times$ 200 grid covering the selected field, and a 2D Gaussian kernel with a full width of $3.5^{\prime}$ was applied to smooth the distribution.

\begin{figure}[h]
\centering
\includegraphics[width=\linewidth]{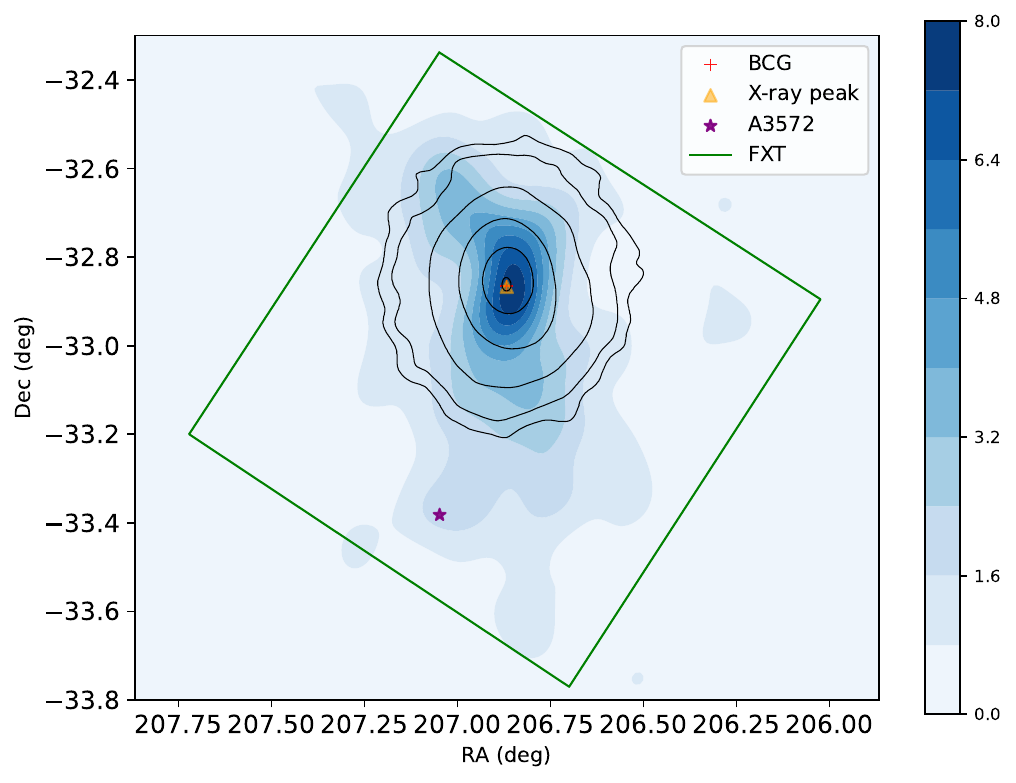}
\caption{Galaxy luminosity density map, generated from optical photometric data (galaxies with $r$-band magnitudes $R_c<22$ and redshifts in the range $0.019$--$0.079$), revealing a north--south extension. The red cross indicates the BCG, the yellow triangle marks the X-ray peak (R.A.\ $=206.86^\circ$, Dec.\ $=-32.86^\circ$), the purple star shows the position of A3572, the green rectangles denote the FoV of \textit{EP-FXT}, and the black contours show the X-ray isophotes.
}
\label{fig:optical}
\end{figure}

The resulting luminosity density map is shown in Fig. \ref{fig:optical}. A prominent central peak (the yellow triangle) is evident, coinciding with the location of the BCG (marked by a red cross) very well. The galaxy distribution exhibits a mild elongation along the north–south direction. The \textit{EP-FXT} field (represented by a green square) fully covers both the core and the extended outskirts of the cluster. 
The overlaid \textit{EP-FXT} X-ray isophotes (shown as black lines) show a more relaxed morphology. Considering that the relaxation time of the gas is shorter than that of the galaxies, this means that this cluster might be a late-stage merger, where the gases has fully merged and the galaxies have not yet relaxed.

There is another galaxy cluster candidate, A3572, located about $30^{\prime}$ to the south of A3571.
\citet{1999Struble} estimated a redshift of 0.052 and a velocity dispersion of 1216 km~s$^{-1}$ based on six galaxies, while \citet{2014ApJlauerBCG} reported a redshift of $z \approx 0.039$ for an elliptical galaxy that is considered to be its BCG.
Based on the photometric luminosity density map and the corresponding X-ray emission distribution, we do not detect any significant enhancement in galaxy number density or X-ray surface brightness at the reported location of A3572.
This suggests that, if A3572 exists, it may host a very weak ICM or be a poor system that is not clearly detectable with the currently available photometric and X-ray data. Consequently, the nature of A3572 remains uncertain.

\section{Analysis and results} \label{sec:3}
The galaxy density distribution of A3571 shows a noticeable elongation along the north–south direction, suggesting that the cluster is not in a state of hydrostatic equilibrium.
To investigate its thermodynamic properties and assess potential merger activity, we conducted a detailed imaging and spectral analysis of the ICM.

\subsection{1D analysis}

\subsubsection{The surface brightness profile}
\label{sec:3.1}

\begin{figure}[h]
\centering
\includegraphics[width=\linewidth]{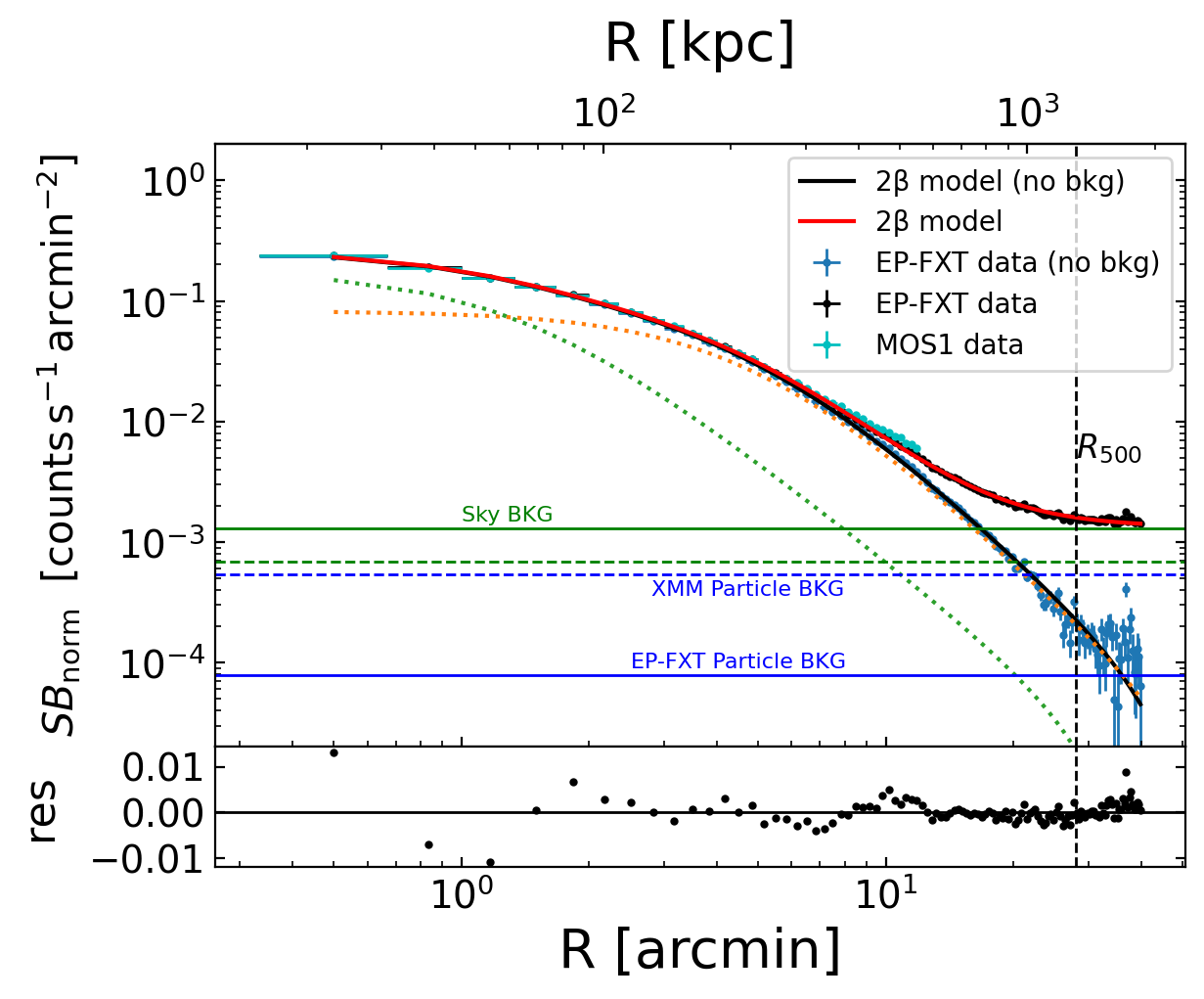}
\caption{Normalized surface brightness profiles measured with \textit{EP-FXT} and \textit{XMM-Newton}. The black points represent the \textit{EP-FXT} data, and the solid red line shows the best-fit double-$\beta$ model, including background components. The blue points correspond to the \textit{EP-FXT} surface brightness after subtraction of both the particle and sky backgrounds, fitted with a double-$\beta$ model without backgrounds. The best-fit curve is shown as the solid black line, while the dashed orange and green lines indicate the two individual components of this model.
The cyan points are the \textit{XMM-Newton} data.
The horizontal solid blue and green lines denote the particle background from \texttt{fxtbkggen} and sky background levels for \textit{EP-FXT}, respectively, while the horizontal dashed blue and green lines indicate the particle and sky background levels for \textit{XMM-Newton}.}
\label{fig:sb_profile}
\end{figure}

To comprehensively assess the background level of \textit{EP-FXT}, we extracted the radial surface brightness profile of A3571, centered on its X-ray emission peak, and compared it with the corresponding profile obtained from \textit{XMM-Newton} observations, as shown in Fig.~\ref{fig:sb_profile}.

Black points are the \textit{EP-FXT} data, which extends out to approximately $40^{\prime}$. 
The horizontal blue line represents the \textit{EP-FXT} particle background $\sim (7.891\pm0.016) \times 10^{-5}$ cts/s/arcmin$^{2}$ , generated using \texttt{fxtbkggen} as described by \citet{2022zhang}. 
The horizontal green line represents the \textit{EP-FXT} sky background $\sim1.29\times10^{-3}~cts/s/arcmin^{2}$, obtained by fitting a constant to the surface brightness in the $30^{\prime}$–$40^{\prime}$ region of A3571, which nevertheless includes faint emission from the cluster.
The blue points indicate the \textit{EP-FXT} surface brightness after subtracting both background components.
We employed a double-$\beta$ model to fit the background-subtracted surface brightness profile. The model equation is
\begin{equation}
S(r)=A[1 + ( \frac{r}{r_{c1}})^2]^{-3\beta_1 + 0.5}+B[1 + ( \frac{r}{r_{c2}})^2]^{-3\beta_2 + 0.5} + C
.\end{equation}

The best-fit parameters are listed in Table~\ref{tab:profiel_sb}.
Additionally, we plot in Fig.~\ref{fig:sb_profile} the two components of the double-$\beta$ model, which reveal the presence of a dense core in the central region of A3571.

Similarly, we performed a radial analysis of the \textit{XMM-Newton} MOS1 data and normalized its surface brightness profile to the maximum value of the \textit{EP-FXT} data. The background components included the sky background obtained from blank-sky observations, and the particle background derived from the 10~–~12 keV spectrum. The dashed blue and green lines correspond to the normalized particle background of $~\sim(5.386~\pm~0.092)~\times10^{-4}~cts/s/arcmin^{2}$ and the normalized sky background of $~\sim~6.880\times10^{-4}~cts/s/arcmin^{2}$, respectively.

As shown in Fig.~\ref{fig:sb_profile}, the surface brightness profiles of EP-FXT and \textit{XMM-Newton} are consistent.  
However, at a radius of around $10^{\prime}$, the \textit{XMM-Newton} surface brightness profile appears slightly higher. This minor difference is likely attributable to increased background uncertainties and reduced effective exposure toward the edge of the \textit{XMM-Newton} FoV. In addition, \textit{EP-FXT} offers a much larger FoV, extending to approximately $40^{\prime}$. 
Additionally, the low particle background of \textit{EP-FXT} 
provides an ideal condition to study diffuse sources.
On the other hand, the sky background of \textit{EP-FXT} is slightly higher than that of \textit{XMM-Newton}, because it is estimated by fitting a constant to the surface brightness in the outer region of A3571, which still contains faint emission from the cluster. In contrast, the \textit{XMM-Newton} background is based on blank-sky observations and is therefore free from cluster contamination.

\begin{table}[h]
\footnotesize
\caption{Best-fit parameters of the double-$\beta$ model with $1\sigma$ errors.}
\label{tab:profiel_sb}
\centering
\begin{tabular}{lc}
\hline
\hline
Parameter & Value \\
\hline
$\beta_{1}$ & $0.723 \pm 0.056$ \\
$\beta_{2}$ & $0.631 \pm 0.221$ \\
$r_{c1}$ & $4.875 \pm 0.561$ \\
$r_{c2}$ & $1.388 \pm 0.303$ \\
$A$ & $0.082 \pm 0.026$ \\
$B$ & $0.176 \pm 0.028$ \\
$C ( 10^{-5}) $ & $-4.222 \pm 9.126$\\

\hline
\end{tabular}
\tablefoot{Best-fit parameters of the double-$\beta$ model with $1\sigma$ errors, based on the background-subtracted. 
$r_{c1}$ and $r_{c2}$ are in units of arcmin. A, B, and C are in units of $cts~s^{-1}arcmin^{-2}$.}
\end{table}

\subsubsection{Global spectral analysis}

\begin{figure}[h]
\centering
\includegraphics[width=0.9\linewidth]{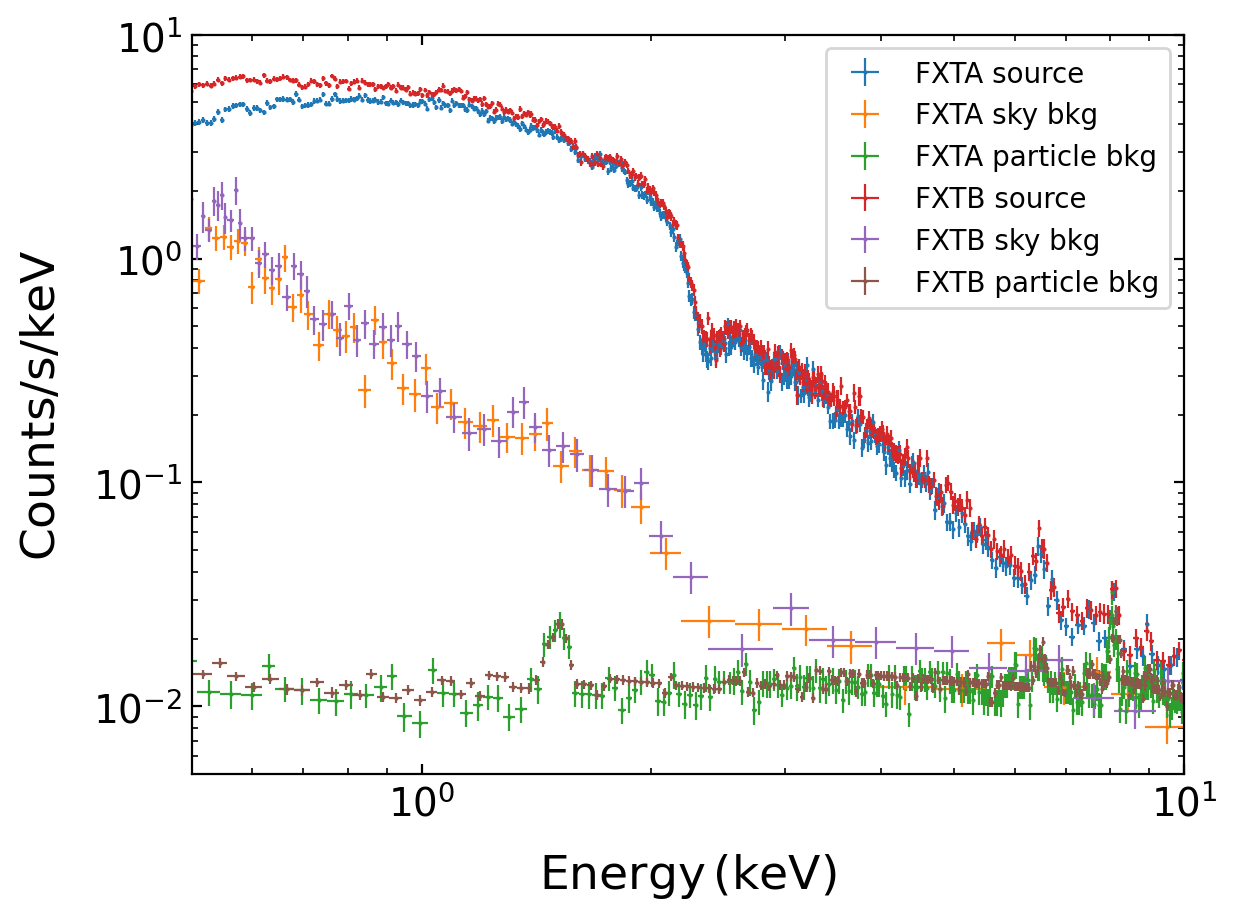}
\caption{Spectrum of A3571 extracted within a $20^{\prime}$ radius centered on the X-ray peak.
The blue and red points represent the observational data from FXT-A and FXT-B, respectively. The green and brown points correspond to their particle backgrounds, and the orange and purple points represent the sky backgrounds after subtraction of the particle background.
}
\label{fig:spectrum}
\end{figure}

In the spectral analysis, we considered both the particle background and the sky background components. The particle background was generated using the \texttt{fxtbkggen} tool, while the sky background was extracted in a circle region beyond $R_{500}$ (see the white circle in Fig.~\ref{fig:image}). 
The source spectrum was obtained from a circular region with a radius of $20^{\prime}$ centered on the X-ray peak (indicated by the red and blue points in Fig.~\ref{fig:spectrum}).
For background, we first subtracted the particle background—scaled to match the sky background region—from the spectrum of the background region shown in Fig.~\ref{fig:image}.
The sky background was then rescaled to match the area of the source region. The sky background and background-subtracted source spectrum are shown as the purple and orange points in Fig.~\ref{fig:spectrum}, respectively.
Similarly, since the particle background generated by \texttt{fxtbkggen} covers the entire CCD, it also needed to be rescaled to match the area of the source region before subtraction. The particle background is shown as the green and brown points in Fig.~\ref{fig:spectrum}.

The particle background spectra of FXT-A and FXT-B exhibit similar shapes, remaining at a level of approximately $10^{-2}$ counts s$^{-1}$ keV$^{-1}$, which is well below the source emission. Three prominent instrumental lines are present: Al-K$\alpha$ ($\sim$1.48 keV), Fe-K$\alpha$ ($\sim$6.38 keV), and Cu-K$\alpha$ ($\sim$8.04 keV).
Around $\sim$1 keV, where the source emission peaks, the particle background contributes less than 1\% of the source count rate, thereby enabling more reliable analysis of faint diffuse sources.
Below 1 keV, the FXT-B spectrum differs from that of FXT-A, primarily due to FXT-B’s larger effective area in this energy band. 
In addition, extended sources are influenced by atmospheric absorption at low energies because of the satellite’s low-Earth orbit. Therefore, we restricted our data processing and analysis to the 1.0–8.0 keV range.

\subsubsection{The radial temperature profile}\label{sec:radial_kt}
\begin{figure}[h]
  \centering
  \begin{minipage}{\linewidth}
    \centering
    \includegraphics[width=0.95\linewidth]{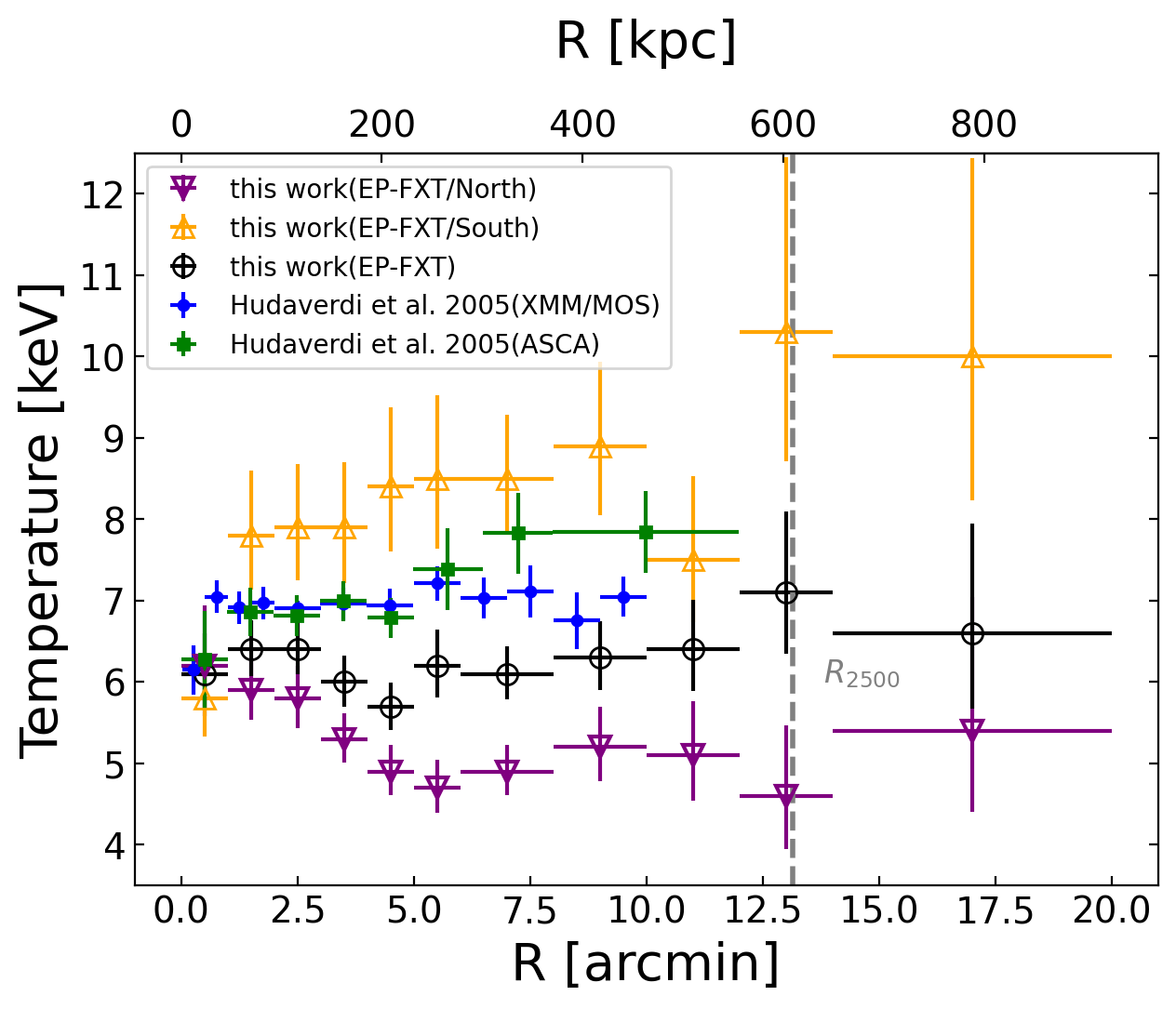}
  \end{minipage}
  \caption{Radial temperature profile of A3571. 
  The black circles show the temperature distribution measured using \textit{EP-FXT} in this study. The blue dots and green squares correspond to the \textit{XMM-Newton} data and the \textit{ASCA} data reported by \citet{2005xmm3571}.
  The purple and yellow triangles indicate the temperature profile measured by \textit{EP-FXT} in the southern and northern directions, respectively.
  The vertical dashed gray line indicates the location of $R_{2500}$.}
  \label{fig:radial_kt_profile}
\end{figure}

Thanks to the large FoV of \textit{EP-FXT}, we were able to   explore the outer ICM of A3571 efficiently.
Centering on the X-ray peak, we used \texttt{XSPEC} (version 12.14.1) to extract spectra from 11 concentric annular regions (more than 20000 photons were detected in each region) in the 1.0–8.0 keV energy band and conducted a radial analysis.
The ancillary response files were generated using the \textit{EP-FXT} response tool \texttt{fxtarfgen}. 
The spectra were grouped using \texttt{grppha} with a minimum of 30 counts per bin to ensure the applicability of the 
$\chi^2$ statistic.
The diffuse ICM thermal emission was modeled using the {\tt APEC} (version 3.0.9), while Galactic hydrogen absorption was accounted for with the {\tt TBABS} model. We adopted the solar abundance table from \citet{aspl2009}, with the redshift fixed at 0.039, the abundance fixed at 0.3 and the hydrogen column density fixed at $5.08 \times 10^{20}~\mathrm{cm}^{-2}$ \citep{willingale2013calibration}. All other parameters were left free during the fit. The radial profiles are shown in  Fig.~\ref{fig:radial_kt_profile}.
The black points represent the \textit{EP-FXT} data in this work, respectively. The blue points indicate the \textit{XMM-Newton} data, while the green points show the \textit{ASCA} data from \citet{2005xmm3571}.

In the central region ($R<1^{\prime}$), the \textit{EP-FXT} temperature is consistent with the results of \citet{2005xmm3571} based on \textit{XMM-Newton} and \textit{ASCA}, thereby reinforcing the evidence for the presence of a cool core. 
Beyond the central region, the temperature profile is broadly consistent with an approximately isothermal distribution at $\sim6$~keV. A mild decrease in temperature is observed around $R \sim 5^{\prime}$, followed by a slight recovery at larger radii.

To further investigate the systematic deviation between temperature of \textit{EP-FXT} and measurements from literature, we divided the region into northern ($0^{\circ}-180^{\circ}$) and southern ($180^{\circ}-360^{\circ}$) sectors with west as $0^{\circ}$ and angles increasing clockwise. The corresponding temperature distributions are shown as the yellow and purple points in Fig. \ref{fig:radial_kt_profile}.

Although the azimuthally averaged temperatures measured by \textit{EP-FXT} are overall lower than those obtained from \textit{XMM-Newton} and \textit{ASCA}, with a difference on the order of $\sim$ 1 keV, the temperature measured by \textit{EP-FXT} in the southern region is higher than in the other two datasets. This behavior may reflect the presence of multiple temperature components within the cluster, to which different instruments respond differently, thereby leading to discrepancies in the derived temperatures; nevertheless, the overall temperature trends remain consistent.

Notably, the northern region also exhibits a clear temperature decline. This feature is associated with the brightness excess detected in the northern region and will be explored further in the following sections.

\subsection{2D analysis}
\subsubsection{The brightness residual image}

\begin{figure}[h]
\centering
\includegraphics[width=\linewidth]{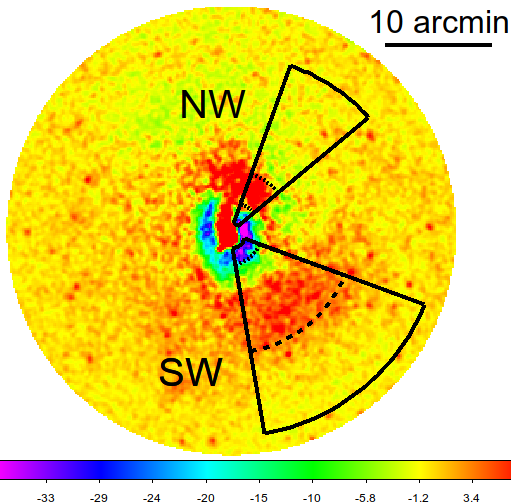}
\caption{Residual map of A3571. It was obtained by subtracting an elliptical $\beta$ model with an added constant component from Fig.~\ref{fig:image}.
The two sectors are regions of radial analysis in the subsequent sections.
}
\label{fig:residual}
\end{figure}

To investigate the spatial structure of A3571, we used \texttt{Sherpa} version~4.16.0\footnote{\url{https://cxc.cfa.harvard.edu/sherpa/}} to subtract an elliptical $\beta$-model, centered on the X-ray peak, together with a constant sky background (see Sect.~\ref{sec:3.1}), from Fig.~\ref{fig:image}. The best-fit parameters of the elliptical $\beta$ model are listed in Table~\ref{tab:ellbeta}, and the residual map is shown in Fig.~\ref{fig:residual}.

A distinct U-shaped surface brightness deficit is found south of the cluster center, whereas a fan-shaped brightness enhancement is detected to the north. 
\citet{2005xmm3571} reported similar features based on \textit{XMM-Newton} observations. 
There is also a faint surface brightness enhancement in the southwestern.
These features suggest that A3571 is not fully relaxed and  still being disturbed.

\begin{table}[h]
\caption{Best-fit parameters of the elliptical $\beta$ model with $1\sigma$ uncertainties.}
\label{tab:ellbeta}
\centering
\footnotesize
\begin{tabular}{lcc}
\hline\hline
Parameter & Value & Unit \\
\hline
$r_{0}$ & $19.40 \pm 0.07$ & pixel \\
Ellipticity & $0.25 \pm 0.001$ & -- \\
$\theta$ & $4.79 \pm 0.003$ & rad \\
Amplitude & $424.43 \pm 1.51$ & counts s$^{-1}$ pix$^{-1}$ \\
$\alpha$ & $1.32 \pm 0.002$ & -- \\
\hline
\end{tabular}
\end{table}

\subsubsection{Thermodynamic 2D maps}\label{sec:FXT_bin_maps}
\begin{figure*}[htp]
    \centering
        \includegraphics[width=0.48\textwidth]{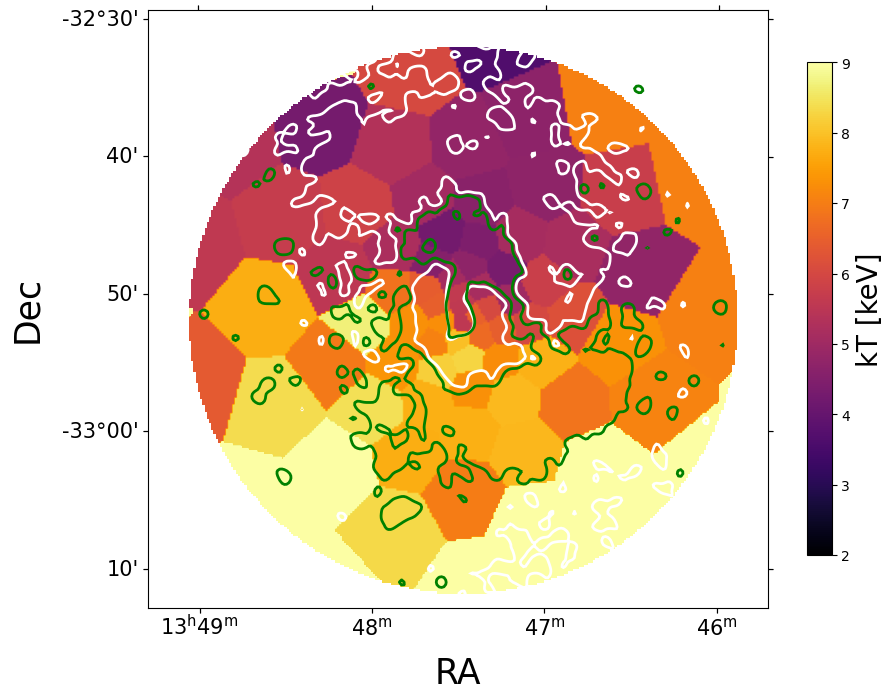}
        \centering
    \hfill
        \includegraphics[width=0.48\textwidth]{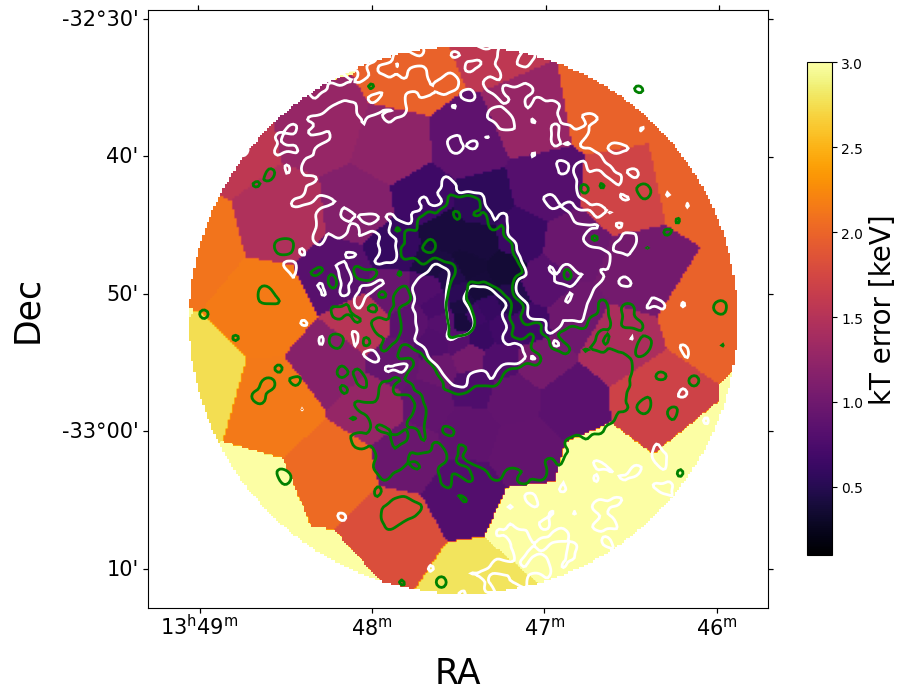}
        \centering

    \vspace{1em}

        \includegraphics[width=0.48\textwidth]{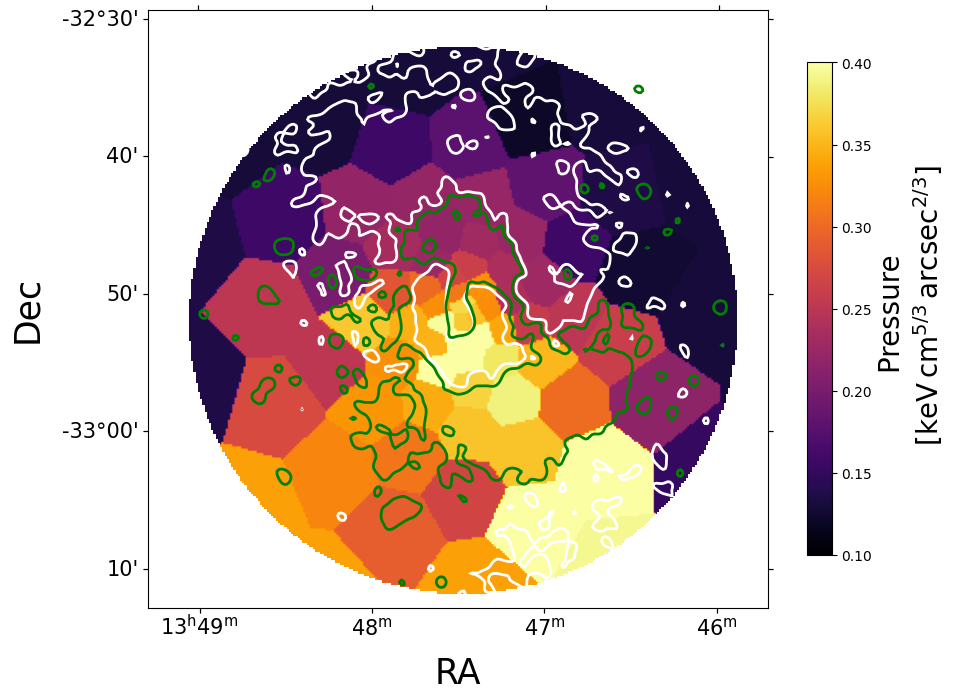}
        \centering
    \hfill
        \includegraphics[width=0.48\textwidth]{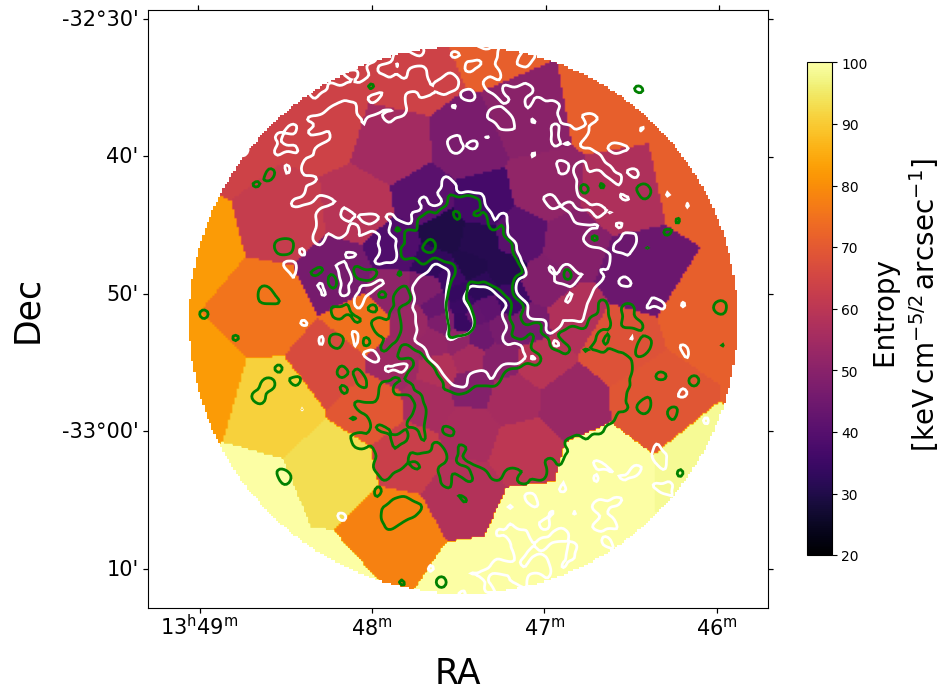}
        \centering
    \caption{Voronoi Tessellation method applied to partition A3571 and spectral fitting performed for each subregion in the 1.0-8.0 keV energy band.
    \textit{Top left}: Temperature map. \textit{Top right}: Temperature error map. \textit{Bottom left}: Pseudo-pressure map. \textit{Bottom right}: Pseudo-entropy map.
    Contours in different colors represent the different intensity levels in Fig.~\ref{fig:residual}.}
    \label{fig:map}
    \end{figure*}

To enable high-resolution 2D mapping of the ICM thermodynamic properties, we performed an Voronoi tessellation on the X-ray image, growing regions to S/N $\ge$ 60 and refining them with S/N‐weighted Lloyd iterations, yielding 102 adaptive bins.
The spectrum extracted from each region was modeled with a {\tt TBABS} * {\tt APEC } model in 1.0-8.0 keV, with the same parameter settings described in Sect.~\ref{sec:radial_kt}.
Based on the best-fitting temperature and normalization values, we derived pseudo-pressure and pseudo-entropy maps using $ P = kTn_{\textnormal{e}}$ and $S = kTn_{\textnormal{e}}^{-2/3}$, respectively \citep{sasaki2016x}, $n_\textnormal{e}$ is electron density and kT is temperature. 
Because the normalization obtained from spectral fitting is proportional to the square of the electron density, we express $P \propto kTnorm^{0.5}$ and $S \propto kTnorm^{-1/3}$ \citep[see Fig. ~\ref{fig:map};][]{2006MNRASfabian}.

The temperature map results indicate that the northern region is overall cooler than the southern region, a trend that is not only consistent with previous \textit{XMM-Newton} studies \citep{2005xmm3571} but also in agreement with the temperature profile shown in Fig. \ref{fig:radial_kt_profile}.
In the overall cooler northern region, the fan-shaped structure shows an even lower temperature, which is consistent with the drop in temperature around $5^{\prime}$ in Fig.~\ref{fig:radial_kt_profile}, supporting the interpretation that this decrease is driven by the fan-shaped structure.
In addition, the surface brightness enhancement found in the southwest coincides with a region of higher temperature.

The pseudo-pressure map clearly indicates that the pressure is lower in the northern region and higher in the southern region. 
In contrast, the pressure distribution derived from \textit{XMM-Newton} observations appears smooth, showing no prominent localized features, and declines steadily with radius in a similar manner along different azimuthal directions \citep{2005xmm3571}.
This difference may arise from the different methodologies used to estimate temperature and density. In the case of \citet{2005xmm3571}, the temperature was estimated from hardness ratios, and the electron density was inferred from the soft-band (0.8–1.6 keV) X-ray surface brightness. While this approach offers advantages in terms of computational efficiency, it may be less capable of resolving subtle structures in dynamically complex regions. In contrast, the \textit{EP-FXT} analysis employs adaptive binning with a S/N of 60 for spectral fitting, yielding higher spatial resolution maps of temperature and normalization, which better reveal localized pressure features.

In the pseudo-entropy map, the northern region displays an overall low entropy distribution. A similar low-entropy feature was also identified by \citet{2005xmm3571}. This indicates that the fan-shaped structure is the disturbed core of the cluster.

At the center, the U-shaped surface brightness deficit exhibits high temperatures and entropy, indicating that it is filled with hot, low-density gas—features reminiscent of X-ray cavities typically driven by active galactic nucleus feedback. However, A3571 lacks the characteristic radio emission commonly observed in such cavities \citep{2002venturi}.
This absence suggests that active galactic nucleus feedback is unlikely to be the dominant cause, and that alternative mechanisms—such as mergers or gas sloshing—offer a more plausible explanation for the origin of the U-shaped structure.

\section{Discussion}
\label{sec:discussion}

Although the X-ray morphology of A3571 looks like relaxed, its
residual map and 2D thermodynamic maps reveal disturbed structure in the inner region. Considering its galaxy density distribution is also elongated along the north–south direction. To identify possible merging signals, 
we selected two sectors covering the brightness excesses in the northwest and southwest for detailed analysis.

\begin{figure}[h]
  \centering
\includegraphics[width=\linewidth]{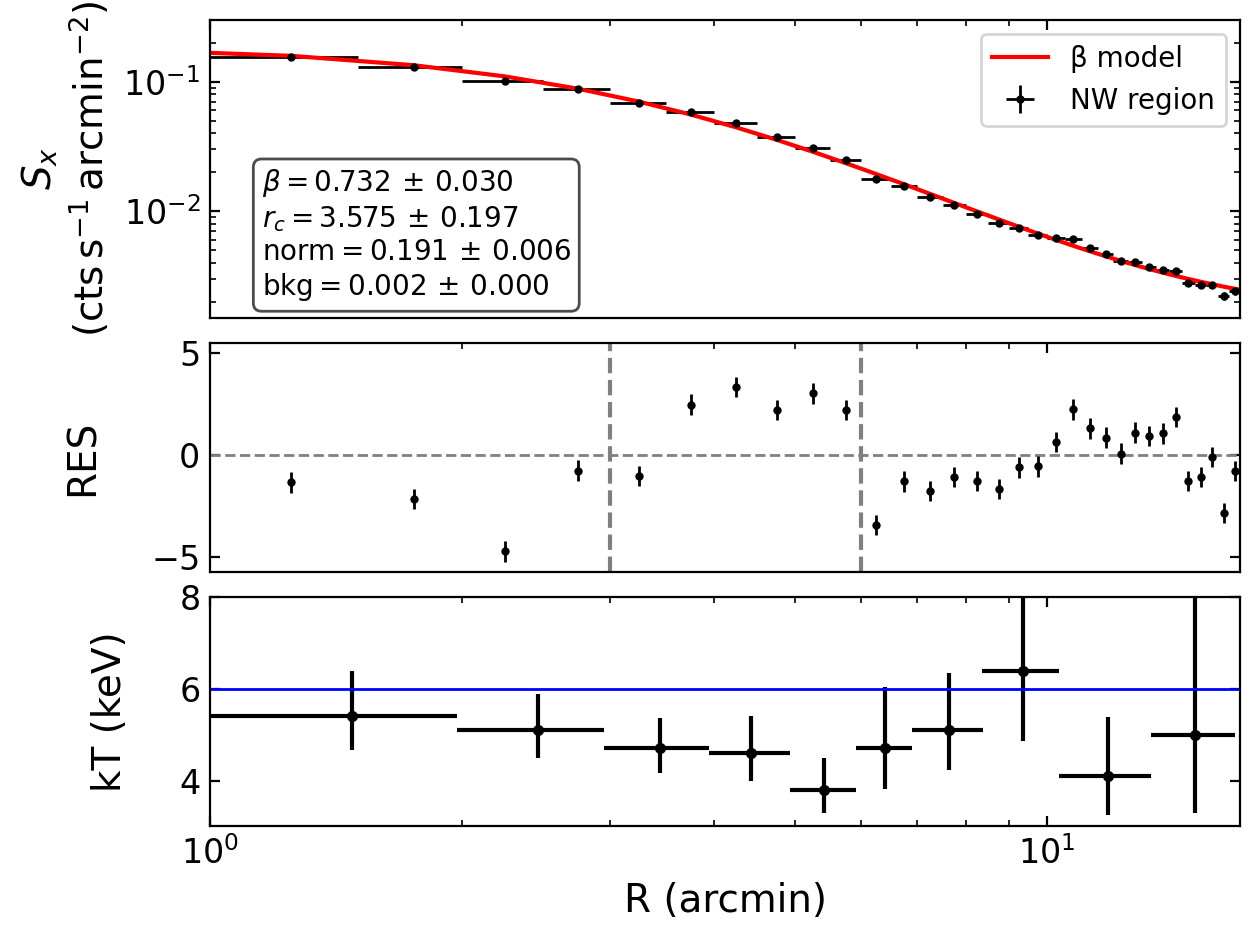}\\[1ex] 
\includegraphics[width=\linewidth]{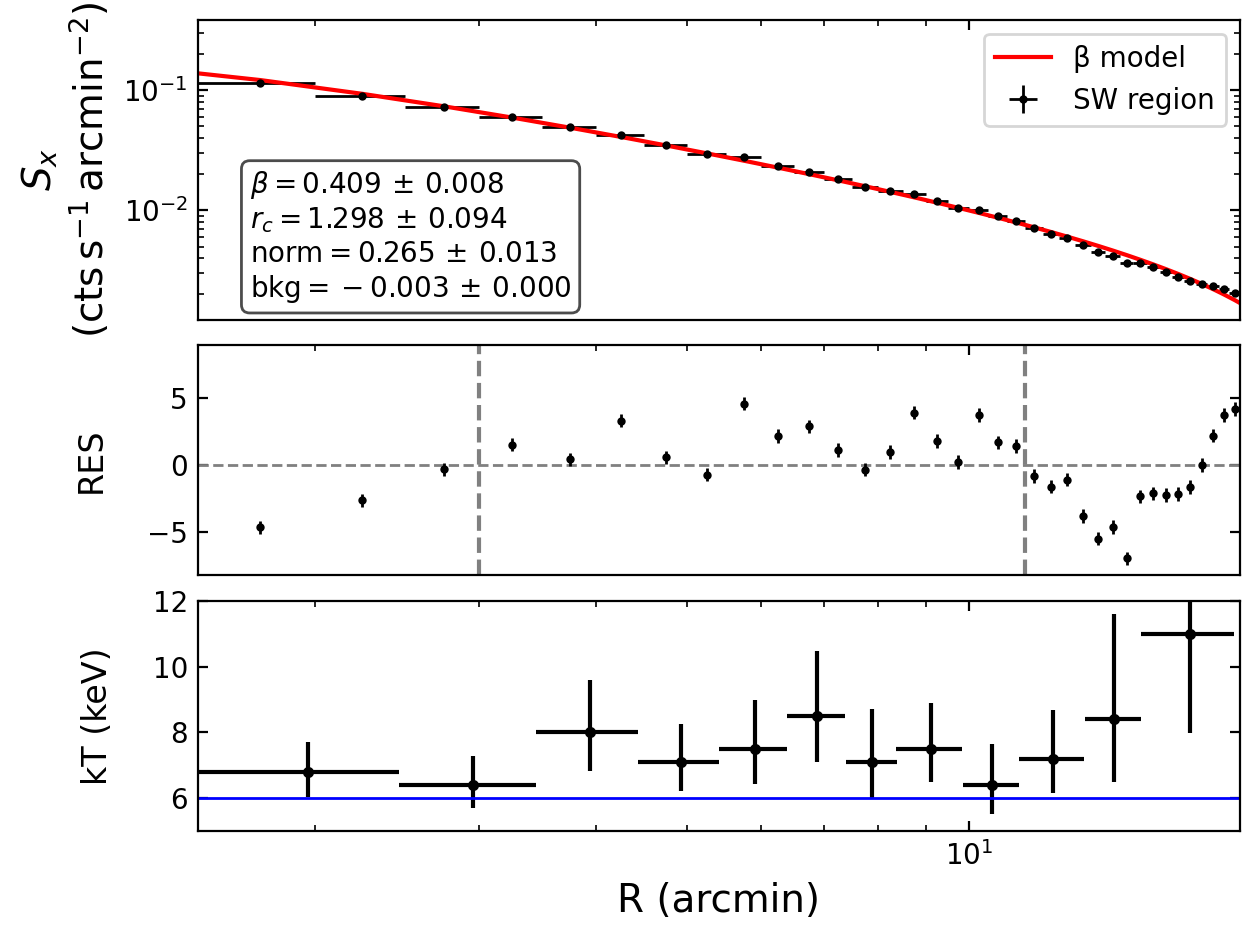}
  \caption{
  Radial temperature and surface brightness profiles in the 1.0–8.0 keV energy band based on the regional divisions shown in Fig.~\ref{fig:residual} (NW and SW). The red curve represents the best-fit model, while the vertical dashed gray lines indicate the locations at which the surface brightness residuals change, corresponding to the dashed black lines in Fig.~\ref{fig:residual}. The horizontal blue line serves as a reference, corresponding to a temperature of 6 keV.}
  \label{fig:radial_profile}
\end{figure}

Figure~\ref{fig:residual} shows the sectors defined in the northwest and southwest directions, from which the radial surface brightness profiles were extracted and fitted using the best-fit model. In addition, spectra were extracted from each sector and fitted with a {\tt TBABS * APEC } model, resulting in radial temperature profiles (see Fig.~\ref{fig:radial_profile}).

The results show that, in the northwestern direction, the radial surface brightness shows a deficiency within a radius of $3^{\prime}$ compared to the model, which corresponds to the outer boundary of the U-shaped structure in that direction. Between $3^{\prime}$ and $6^{\prime}$, the surface brightness shows an excess corresponding to the fan-shaped structure in the residual map.  
Around $6^{\prime}$, the radial temperature profile shows no pronounced discontinuity and exhibits only a slight increase.
However, the line-of-sight projection of the cold sloshing tail could reduce the apparent temperature contrast.
Alternatively, the presence of a weak cold front at this radius cannot be excluded.

In the southwestern direction, the surface brightness residual exhibits a deficiency within a radius of $3^{\prime}$, while at approximately $11.5^{\prime}$, the brightness shifts from an excess to a deficiency. This location may correspond to the outer boundary of the enhanced-brightness region in the southwest. However, within this region, the overall temperature remains consistently above 6~keV, but the radial temperature profile shows no pronounced discontinuity; consequently, there is no evidence of shock heating. 

By combining the 1D~(see Fig.~\ref{fig:radial_kt_profile}) and 2D (see Fig.~\ref{fig:map}) temperature analyses, it is clear that the northern region is significantly cooler than the south. In Fig. \ref{fig:radial_kt_profile}, the temperature drop around $5^{\prime}$ is primarily associated with the northern sector, which corresponds in Fig. \ref{fig:map} to a fan-shaped region of low temperature and low entropy. In contrast, the \textit{XMM-Newton} results do not show a pronounced temperature decrease, likely due to the limited FoV and reduced sensitivity. Nevertheless, the \textit{XMM-Newton} data also reveal a temperature asymmetry between the north and south. 
We therefore suggest that this feature likely results from dynamical activity along the north–south axis.

Based on the above analysis, we propose that the structure of A3571 originates from gas sloshing triggered by the off-axis passage of a low-mass subcluster moving from south to north. The sloshing displaces low-entropy gas from the cool core, producing a fan-shaped brightness excess to the north. At the same time, the central region, originally filled with dense, cool gas, is replenished with diffuse hot gas, resulting in the observed U-shaped surface brightness deficit.
This is consistent with the results of numerical simulations \citep{Ascasibar2006}. 
These simulations show that even a low-mass subcluster, during an off-axis passage, can gravitationally disturb the main cluster, shifting its central mass peak and, through a reversal of ram pressure, eject cool gas from the bottom of the potential well. This process initiates persistent sloshing that can naturally generate fan-shaped distributions of cool gas and U-shaped surface brightness deficits similar to those observed in A3571.

\section{Conclusion}
\label{sec:conculsion}

We performed a systematic X-ray analysis of the galaxy cluster A3571, combining optical observations with X-ray data obtained from the \textit{EP-FXT} telescope, which features a large FoV and low particle background. We measured the cluster’s outer properties beyond $R_{500}$ in the imaging domain.
The main results of this work are summarized as follows:

\begin{enumerate} 

\item Our data provide a high-quality image of A3571 within radius $R_{500}$. It exhibits a relatively regular morphology. 
Both the surface brightness and temperature profiles exhibit typical cool-core characteristics.

\item Two prominent regions of surface brightness excesses are detected within $20^{\prime}$ of the cluster center, located to the north and southwest. The southwestern excess is associated with high-temperature features, whereas the northern excess corresponds to a cooler structure. The northern low-temperature feature is most likely produced by gas sloshing occurring primarily along the line of sight and triggered by the off-axis passage of a low-mass subcluster.

\item The temperature distribution of A3571 shows a pronounced north–south asymmetry, consistent with previous \textit{XMM-Newton} observations. A similar north–south elongation is also evident in the optical galaxy density distribution, suggesting that the cluster’s merger activity occurred along this axis.

\item We detect no thermal emission from the direction of A3572, nor do we find any excess in the density of optical galaxies. This implies that, if A3572 exists, it may host an extremely weak ICM or represent a poor system that is not detectable with the currently available X-ray and optical data.
In contrast, the galaxy density distribution and X-ray morphology indicate that A3571 is still recovering from a minor merger and is currently in the post-merger phase.

\end{enumerate}

The results of this study are limited by the calibration uncertainties of \textit{EP-FXT}, which primarily affect measurements of the outer temperature and thermodynamic properties. Data from the formal observational phase (beginning in June 2024) are not affected by this issue. We expect that more comprehensive studies of the hot gas in galaxy clusters will be possible with these improved data.

\begin{acknowledgements}
      This work has been supported by the National Natural Science Foundation of China No. 12573003, the National Key Research and Development Program of China (No. 2023YFC2206704), National Key R\&D Program of China No. 2025YFF0511104, the International Partnership Program of Chinese Academy of Sciences, Grant No. 013GJHZ2024015FN and the China Manned Space Program with grant No. CMS-CSST-2025-A04. 
      A.L. acknowledges the supports from the National Natural Science Foundation of China (Grant No. 12588202). 
      This work is based on data obtained with \textit{Einstein} Probe, a space mission
    supported by Strategic Priority Program on Space Science of Chinese
    Academy of Sciences, in collaboration with ESA, MPE and CNES (Grant
    No. XDA15310000).
    EP is a space mission supported by
Strategic Priority Program on Space Science of Chinese Academy
of Sciences, in collaboration with ESA, MPE and CNES (Grant
No.XDA15310303, No.XDA15310103, No.XDA15052100).
\end{acknowledgements}

\bibliographystyle{aa}
\bibliography{ref}

\appendix
\section{Comparison of temperatures}\label{app:fxt}

\begin{figure}[h]
  \centering
\includegraphics[width=0.9\linewidth]{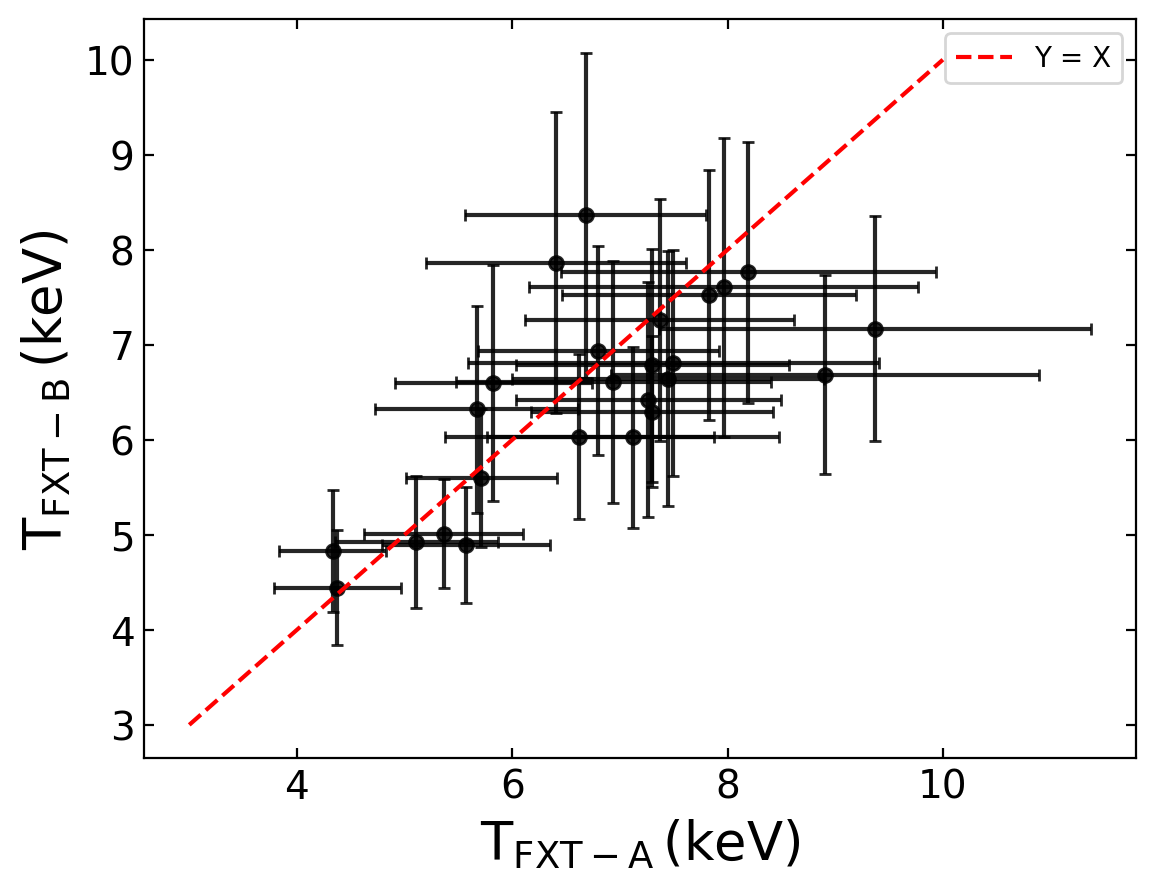}\\[1ex] 
  \caption{Comparison of FXT-A and FXT-B temperatures in the
1.0-8.0 keV energy band. The dashed red line represents 1:1.
  }
  \label{fig:compare_kt}
\end{figure}

To evaluate the consistency between FXT-A and FXT-B, we performed spectral fits in the 1.0–8.0 keV band within the Voronoi-defined regions (see Sect.\ref{sec:FXT_bin_maps}), using the \texttt{TBABS*APEC} model with the same parameter settings as in Sect.\ref{sec:radial_kt}. The temperature measurements were then compared (see Fig.~\ref{fig:compare_kt}). The results show that the temperatures obtained with FXT-A and FXT-B are consistent within the uncertainties. Although FXT-A yields slightly higher values than FXT-B, this discrepancy is likely due to FXT-B’s larger effective area at low energies and calibration differences between the two instruments.

In addition, the observation of A3571 was carried out during the performance verification phase in March, whereas the formal calibration campaign was not systematically initiated until April.
Consequently, in the absence of finalized calibration products, these data are unavoidably affected by calibration-related systematic uncertainties.

\end{document}